# Delayed photoisomerisation of the trans-PSB3 retinal toy model using on-the-fly quantum dynamics[†]

María Mallo,[a] Susana Gómez-Carrasco,[a] and Sandra Gómez[b]*

We explore the trans-cis photoisomerisation process in a representative retinal protonated Schiff base known as trans-PSB3, employing the quantum dynamics method direct dynamics variational multiconfigurational gaussian -DD-vMCG- in full dimensionality, i.e., 36 degrees of freedom on potential energy surfaces computed *on-the-fly* using the SA(2)-CAS(6,6)SCF electronic structure method with the 6-31G basis set. Although the toy molecule has been extensively studied using trajectory methods such as Tully Surface Hopping and Ab Initio Multiple Spawning, the *on-the-fly* quantum dynamics method DD-vMCG shows a trans-cis isomerisation hundreds of femtoseconds slower using the same electronic structure method, which can be explained in terms of the accesibility to the conical intersection connecting the ground and the excited state.

## 1 Introduction

Vision in humans begins when light enters the eye and reaches the retina, where photoreceptor cells (rods for low-light vision and cones for color vision) convert light into neural signals interpreted by the brain.[1,2] In rods, the primary photoreceptor is rhodopsin, a complex of the opsin protein and the 11-cis-retinal chromophore, linked by a protonated Schiff base (PSB).[3] Upon photon absorption, the retinal chromophore undergoes a rapid photoisomerisation from the 11-cis to the all-trans configuration,[4,5] triggering a cascade of biochemical reactions that culminate in optic nerve stimulation and vision.

This cis-trans photoisomerisation occurs on an ultrafast timescale (within 200 fs) and with high quantum efficiency,[6,7] proceeding through a conical intersection between the first excited ($\pi\pi^*$) and ground electronic states.[8] The mechanism has been extensively investigated through both experimental and computational studies,[9–12], focusing on the excited state dynamics of the chromophores and their role in vision.

Given the complexity of the full rhodopsin system, simplified models such as the penta-2,4-dieniminium cation (PSB3) have become invaluable for probing the fundamental photochemistry of retinal.[13–15] PSB3 preserves the essential features of the retinal protonated Schiff base, including the central double bond and isomerisation pathway, while its small size enables high-level quantum chemical calculations of excited state potential energy surfaces.[16]

Nonadiabatic molecular dynamics (NAMD) simulations are crucial for studying photoinduced processes, but the diversity of available methods presents challenges in accuracy and reproducibility. The need for reliable benchmarks has led to the widespread adoption of PSB3 as a model system for testing and validating NAMD approaches.[17] Its well-characterized photochemistry and reduced complexity have facilitated the development of a plethora of computational results, supporting method refinement and reproducibility, and helping theoretical advances to be extended to more complex systems.

Early computational studies mapped the potential energy surfaces (PESs) of PSB3 and identified the conical intersections (CIs) involved in its cis-trans isomerisation, using high-level methods such as CASPT2 and MRPT2.[18–22] Further work compared quantum chemical methods (e.g., CASPT2, CASPT3, ADC(2)), highlighting the influence of methodological choices on predicted excitation energies and CI branching spaces.[23,24] These studies highlighted the computational challenges of accurately modeling excited state processes, even in minimal systems.

*On-the-fly* nonadiabatic dynamics, employing techniques such as Ab Initio Multiple Spawning (AIMS) and surface hopping, have further explored the role of dynamic nuclear effects and nonadiabatic coupling in PSB3 photoisomerisation.[25–27] These investigations revealed that relevant conical intersections often lie far from minimum energy regions, with vibrational modes driving the isomerisation. Additional approaches, including exact factorization methods[9,28,29] and machine learning-enhanced molecular dynamics,[30] have also provided insight into the mechanisms and global reaction coordinates of retinal models.

Despite lacking some structural features of the full retinal system, PSB3 ($C_5H_6NH_2^+$, see Figure 1) is one of the simplest minimal models for 11-cis-retinylidene,[31] capturing key aspects of the isomerisation mechanism and enabling qualitative understanding of retinal photochemistry.[9,32,33] Its suitability for *ab initio* calculations involving the ground and first excited electronic states makes it an ideal benchmark system.

In this work, we employ for the first time the direct dynamics variational multiconfigurational Gaussian (DD-vMCG) method to study the photoisomerisation of trans-PSB3. This approach will allow us to accurately capture nuclear dynamics and probe fundamental aspects of rhodopsin photochemistry, as well as to compare the performance of this relatively new nonadiabatic dynamics method in predicting the isomerisation.

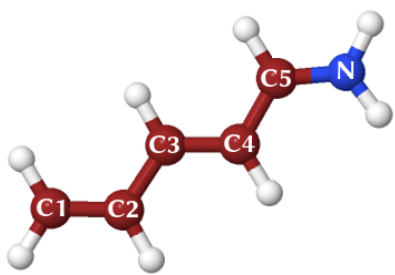

**Fig. 1** The PSB3 retinal model system: 2-trans-penta-2,4-dieniminium cation.

## 2 Computational details

### 2.1 Electronic structure

The electronic structure calculations were performed using a two state average complete active space self-consistent field theory (SA(2)-CASSCF) with an active space consisting of six electrons in six orbitals, denoted as CAS(6,6). This active space, shown in Figure 2, includes three pairs of $\pi$ and $\pi^*$ orbitals, capturing the key electronic configurations responsible for the photoisomerisation process in PSB3. The state specific approach SS-CASSCF with the same active space was used to optimise the ground state minimum and calculate its vibrational frequencies (geometric parameters and frequencies shown in Tables S1 and S5 in the supporting information). The state-averaged approach (SA(2)-CASSCF(6,6)) was applied over the two lowest singlet electronic states, $S_0$ (ground state) and $S_1$ (excited state). In terms of molecular orbitals, $S_0$ and $S_1$ states are characterised by $\pi^2$ and $\pi\pi^*$ electronic configurations, respectively. SA(3)-CASSCF(6,6) calculations were attempted to also include the effects of the second excited state ($S_2$), leading to convergence problems. As dynamics studies reveal that the $S_2$ state is not involved in the population transfer,[25] only two-state average calculations were finally performed. The 6-31G basis set was selected for its ability to significantly reduce computational cost while still accurately reproducing vertical excitation energies, conical intersection locations and energies, and reaction paths when compared with polarized basis sets.[9,32,34]

Previous works, such as those by Martínez et al.[25] Szymczak et al.[32] and Filatov et al.[26], have employed the same CASSCF active space and the 6-31G basis set for studying nonadiabatic dynamics in model retinal chromophores. In the study by Martínez' group, a three-state averaging procedure was adopted,[25] whereas in our work, we focus on a two-state average, emphasizing the dynamics between the ground $S_0$ and the first excited state $S_1$. Szymczak et al.[32] also employed SA(2)-CASSCF(6,6) for the initial reference wavefunctions in their study, followed by multireference configuration interaction (MRCI) calculations. However, our approach remains within the SA(2)-CASSCF framework without further post-CASSCF correlation treatments, as we aim to compare the performance of the *on-the-fly* DD-vMCG method, that would otherwise be too expensive to use.

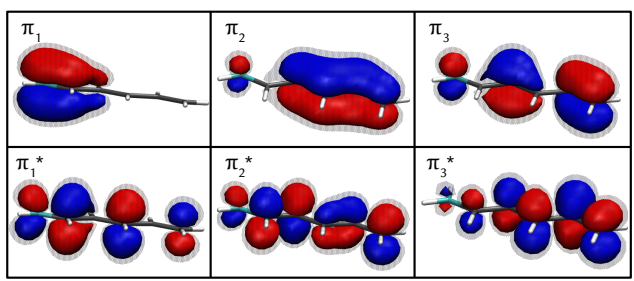

**Fig. 2** Orbitals included in the active space for the SA(2)-CAS(6,6)SCF/6-31G electronic structure calculations. The $S_1$, with $\pi\pi^*$, corresponds to an electron excited from orbital $\pi_3$ to $\pi_1^*$

Table 1 shows the ab initio optimised and vertical excitation energies (in eV) of both electronic states ($S_0$ and $S_1$) for the trans, the cis-PSB3 isomer and the minimum energy conical intersection(MECI_opt) using a SA(2)-CAS(6,6)SCF/6-31G method with the Molpro electronic structure program [35]. MECI_db corresponds to the PSB3 geometry with the lowest energy gap encountered in the dynamics.

|  | $\pi^2/S_0$ | $\pi\pi^*/S_1$ | |
| --- | --- | --- | --- |
|  | optimised | vertical | optimised |
| Trans-PSB3 | 0.00 | 4.84 | 4.45 |
| Cis-PSB3 | 0.13 | 4.69 | 4.53 |
| MECI_opt ($S_0/S_1$) | 2.70 | - | 2.70 |
| MECI_db ($S_0/S_1$) | 2.61 | - | 2.62 |

**Table 1** *Ab initio* optimised and vertical excitation energies (in eV) of both electronic states ($S_0$ and $S_1$) for the most relevant geometries marked in light pink in Figure 4: the Trans, the Cis PSB3 isomer and the MECI_opt point. Calculations are done with a SA(2)-CAS(6,6)SCF/6-31G method using the Molpro program [35]. Also the MECI_db, which is the most accesible point in the database with the lowest energy gap encountered in the dynamics, is given.

### 2.2 The vMCG quantum dynamics method

In the variational Multi-Configurational Gaussian (vMCG) method, the nuclear wavefunction is expressed as a linear combination of time-dependent Gaussian basis functions (GBFs), avoiding the need for precomputing potential energy surfaces on a full primitive grid. This approach belongs to the family of Gaussian wave packet (GWP) methods, where the system's degrees of freedom are described by multidimensional frozen Gaussians, making it well-suited for nonadiabatic quantum dynamics simulations. [36]

The vMCG wavefunction ansatz takes the following form:

$$\Psi(q,t) = \sum_i A_i(t) G_i(q,t) |s\rangle \quad (1)$$

Here, $G_i$ represents separable, frozen GBFs, which are products of one-dimensional Gaussian functions for each coordinate q and $|s\rangle$ is a vector representing the associated electronic state basis. These GBFs are parameterized by time-dependent quantities such as position, momentum, and width. As the GBFs evolve over time, they provide an accurate description of the system's nuclear wavefunction. The method employs a variational principle to determine the positions of these Gaussians over time, keeping the nuclear basis set size optimally small. [37] This distinguishes vMCG from other methods like full multiple spawning (FMS) [38] and coupled coherent states, [39] which rely on classical trajectories for Gaussian centers.

The time-dependent Schrödinger equation (TDSE) is solved using the vMCG ansatz (equation 3) and the Dirac-Frenkel variational principle: [40]

$$\langle \delta\Psi | \hat{H} - i\frac{\partial}{\partial t} | \Psi \rangle = 0 \quad (2)$$

This principle leads to coupled equations of motion for the expansion coefficients $A_i^{(s)}$ and the GBF parameters, resulting in quantum "trajectories" for the GBF centers. These "trajectories" differ from classical ones, as the GBFs are variationally coupled, which allows the method to capture quantum coherence and avoid the need for large basis sets typical in trajectory-based methods. If the Hamiltonian matrix elements $<G_i|H|G_j>$ are calculated exactly, the vMCG method can, in principle, converge to the numerically exact solution of the TDSE.[37] In practice, however, we do not possess information of the full potential energy surfaces and we use the Local Harmonic Approximation (LHA) to evaluate the matrix elements. The potential V(q) is expanded around the center of each GBF ( $G_i(q_0)$ ) up to second order:

$$V(q) = V(q_0) + \frac{\partial V(q_0)}{\partial q}(q-q_0) + \frac{1}{2},\frac{\partial^2 V(q_0)}{\partial^2 q}(q-q_0)^2 \qquad (3)$$

, as if each gaussian were feeling its own quadratic vibronic coupling potential. The zeroth, first and second order energy terms are calculated *on-the-fly* as electronic energies, gradients, nonadiabatic couplings and hessians using *ab initio* quantum chemistry software. This avoids the need for precomputed potential energy surfaces, while maintaining analytical potential form for the integrals over the conformational space. The DD-vMCG (where DD stands for Direct Dynamics, meaning *on-the-fly*) method integrates these features by dynamically calculating the potential energy surfaces as the nuclear wavefunction evolves, making the approach computationally feasible for complex molecular systems.[41]

In DD-vMCG, the wavefunction is propagated on smooth diabatic potential energy surfaces. The propagation diabatisation technique[42] is used to construct global diabatic states from adiabatic ones along the nuclear trajectories, which is critical to avoid the geometric phase issues associated with the adiabatic picture, particularly near conical intersections.[43] New electronic structure points, including energies, gradients, Hessians, and nonadiabatic couplings, are calculated *on-the-fly* and stored in a database. If a new geometry is encountered, it is compared to stored points, and interpolation between the diabatic energies is performed using Shepard interpolation.[44] To further reduce computational cost, the Hessian, one of the most computationally expensive quantities to calculate, is calculated at the initial structure and then approximated during the dynamics using a Hessian updating scheme based on gradient information.[45,46] This practice, combined with avoiding redundant quantum chemistry calculations via database interpolation, dramatically improves the efficiency of the method while maintaining accuracy.

The DD-vMCG method is implemented within the Quantics package developed in the Worth group.[47] The simulation begins by projecting the nuclear ground state onto the first excited state, which initializes a single Gaussian with amplitude, while the other gaussians in the expansion have zero initial amplitude but are displaced in momentum space to prepare the system for propagation. Thanks to the variational nature of the equations of motion, unpopulated Gaussians evolve and contribute to the dynamics without the need for explicit sampling as in trajectory-based methods. This allows the method to optimally describe the evolving wavepacket during the dynamics.

## 3 Dynamic results

Initially, DD-vMCG quantum dynamics propagations were performed using four gaussian basis functions (GBFs). These preliminary simulations generated a small database of electronic structure points, which served as the foundation for subsequent simulations with an increasing number of GBFs. As the wavepacket explored new regions of the potential energy surface, additional points were added to the database, gradually expanding its coverage.

In the final set of DD-vMCG simulations, the dynamics was propagated for up to 300 fs. The largest database, which contained 48,857 electronic structure points, was constructed incrementally by adding points from previous simulations. Specifically, the database for simulations with 8, 16, and 20 GBFs was built upon the earlier 4 GBF simulation, with each successive simulation expanding the database created by the preceding one. This strategy ensured that the wavepacket had access to an increasingly comprehensive representation of the PES as the number of GBFs grew.

The key outputs analyzed in this study are: diabatic electronic populations of the two involved electronic states, one-dimensional cuts of the PES connecting the Franck-Condon region to the conical intersection, the geometries of the database points and the time evolution of the expectation values for every degree of freedom (in the Supplementary Information). Additionally, the trans-cis photoisomerisation process was characterized by monitoring the evolution of the dihedral angle and the bond length alternation parameter, focusing on the weighted average of the Gaussian centers to capture the isomerisation pathway.

## 3.1 Evolution of state populations and geometrical observables

The DD-vMCG dynamics simulations provide access to the full evolution of the gaussian wave packet. This GWP is represented as a linear combination of frozen multiconfigurational GBFs. Since these GBFs are expanded in the nuclear basis of the molecule's normal modes, we can directly extract the expectation values and uncertainties of these modes. At each time step, we determine whether the Gaussian center is close to any previously calculated *ab initio* points stored in the database. This involves applying a unitary transformation from normal mode coordinates to Cartesian coordinates. Using the Cartesian geometries of the GBF centers and their weight in the frozen gaussian expansion, we calculate the average torsional angle as

$$\langle x \rangle(t) = \sum_{i=1}^{N} \text{GGP}_i(t) \cdot x_i(t) \quad (4)$$

and its time dependent standard deviation as

$$\text{StdDev}(t) = \sqrt{\sum_{i=1}^{N} \text{GGP}_i(t) \cdot x_i^2(t) - \langle x \rangle^2(t)} \quad (5)$$

where $\langle x \rangle(t)$ represents the time-dependent average of the torsional angles among atoms C2, C3, C4 and C5 (line marked as "Average" in Figure3. ), StdDev (t) represents the time-dependent standard deviation, GGP are the gaussian gross populations -the square of the $A_i$ coefficients in equation 1-, and N is the total number of gaussians (20 for the plots in Figure 3 and 16 for Figure 6).

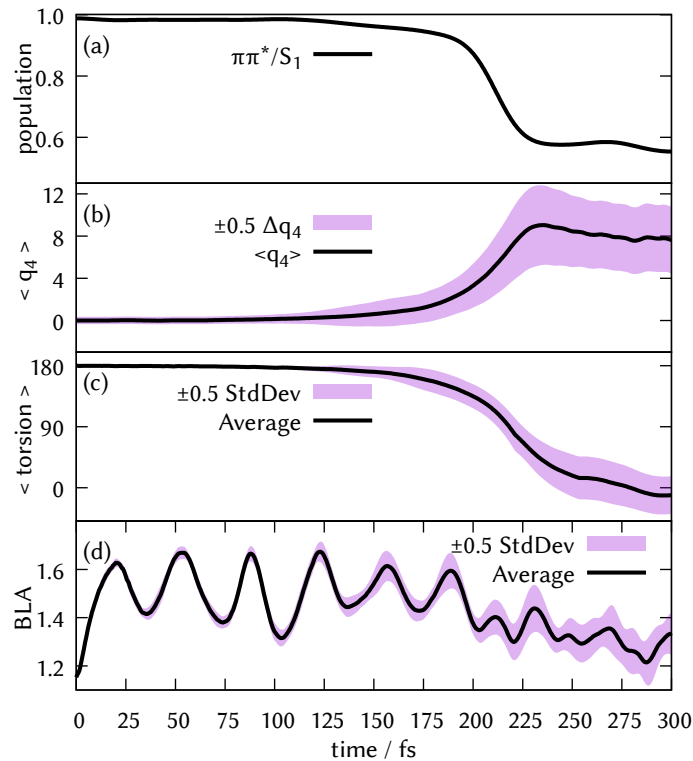

**Fig. 3** (a) Diabatic electronic population decay for the initially excited state ($\pi\pi^*$), (b) Evolution of expectation value and uncertainty (standard deviation) of torsional normal mode $q_4$, (c) average torsional angle (C2-C3-C4-C5) in degrees and standard deviation of the gaussian centers in cartesian coordinates, (d) Evolution of bond length alternation (BLA) parameter in Å defined as $r_{C1-C2} - r_{C2-C3} + r_{C3-C4} - r_{C4-C5} + r_{C5-N}$. For DD-vMCG calculations with 20 GBFs.

The bond-length alternation (BLA) parameter has been calculated following a similar approach. First, the BLA value for each Gaussian center was determined as $x_i(t) = r_{C1-C2} - r_{C2-C3} + r_{C3-C4} - r_{C4-C5} + r_{C5-N}$ where $r_{ij}$ represents the bond lengths between the respective atoms. This value was then weighted by the Gaussian gross population associated with each center.

Figure 3(a) illustrates the ability of the DD-vMCG method to capture the population transfer from the $S_1$ state to the $S_0$ state, which decreases from 100% to 60% over 250 fs. At first glance, this might suggest that isomerisation is incomplete within this timescale. However, insets (b) and (c) of Figure 3 show the expectation value of the torsional normal mode and the average torsional angle, respectively. These data confirm that by 250–300 fs, the PSB3 molecule has fully isomerised from the trans to the cis form, as the torsional angle transitions from approximately 180 degree to 0.

Although the torsional motion is inactive prior to 150 fs, Figure 3(d) shows that bond lengths begin to stretch and elongate within the first 25 fs. During this period, the bond-length alternation (BLA) parameter increases from values below 1.2 -characteristic of the trans-isomer- to almost 1.6, corresponding to the $S_1$ minimum. By 200 fs, the system stabilises, oscillating around a BLA value of 1.4, which corresponds to the cis-isomer. The narrow standard deviations indicate that the isomerisation is effectively complete within this timeframe.

When examining the time evolution of the normal modes of PSB3 (Figures S6–S11), the system can be divided into tuning modes, which carry the gradient and move the wavepacket away from the Franck–Condon region, and coupling modes, which couple different electronic states, facilitate population transfer, and modulate the width of the wavepacket. By analyzing how the expectation value and standard deviation of each normal mode change over time, it is possible to track the most active modes during the dynamics.

During the first 30 fs, the dynamics is primarily driven by mode v27 (tuning) and mode v36 (coupling), which elongate and shorten the central CC bonds. Between 30 and 100 fs, the dynamics become more complex, involving additional tuning modes (v6, v16, and v20) and coupling modes (v16, v21, v23, and v34). These modes predominantly consist of CC stretching motions but gradually incorporate torsional motions, such as mode v16, which corresponds to a slight $NH_2$ torsion. The normal modes of the PSB3 cation are depicted in Figure S5 and Table S5.

From 100 to 200 fs, the previously mentioned tuning modes continue to play a significant role, but at 115 fs, the main torsional mode, v4, emerges as a coupling mode (depicted in Figure 3(b)). This mode regulates the width of the wavepacket and facilitates moderate population transfer. Between 200 and 210 fs, mode v4 becomes the dominant tuning mode, exhibiting the largest amplitude motion, while v23 and v19 govern the coupling between electronic states.

After 220 fs and continuing until the end of the dynamics (300 fs), modes v32 and v30 (CH stretches) display the largest amplitudes, likely driving the system toward the cis structure minimum.

### 3.2 Geometries of the database points

Since the DD-vMCG method stores all *ab initio* and interpolated points in a database containing geometries, energies, gradients, couplings, normal mode displacements, and Hessians we can use these data to represent the conformational space explored during the dynamics. In Figure 4, we depict the torsional angles and BLA parameters of the 48,857 database structures, represented as a 2D histogram with 30 bins per dimension (approximately 6 degree of torsion per bin and 0.07 angstrom per BLA bin). The color of each bin indicates the frequency of data points within that bin, with the darkest color corresponding to bins containing 400 or more structures. The crosses in the plot represent key geometries: the initial trans-PSB3 structure (S0_trans), the final cis-PSB3 structure (S0_cis) on the ground electronic state and the minimun energy conical intersection (MECI_db), which corresponds to the structure with the lowest energy gap between the two electronic states. These three structures are found in our database during the DD-vMCG/20GBFs dynamical simulation. The figure also shows the *ab initio* optimised trans-PSB3 (S1_trans), cis-PSB3 (S1_cis) structures for the excited electronic states and minimum energy conical intersection (MECI_opt). Optimised structures were calculated using a SA(2)-CAS(6,6)SCF/6-31G with Molpro[35] and their energies are given in Table 1.

Notably, around the top right yellow cross representing the minimum of the excited state for the trans-PSB3 geometry, we observe a high density of structures with increasing BLA values, consistent with the dynamical trends shown in Figure 3(d). The MECI_db identified in our simulation lies near the center of the most frequently visited regions of the database. Furthermore, the high density of structures near 0 degree torsion and a BLA value of 1.4 Å indicates successful isomerisation to the cis-PSB3 geometry. Another prominent region in the conformational space corresponds to around 60 degrees of torsion and a BLA of 1.4 Å, which lies close to the optimised MECI geometry, MECI_opt. We further analysed the number of database points that lay around the yellow crosses on the map (using an interval of ± 3% BLA and torsion values of the cross geometries) and found that a 3.6% of the structures lay around the S1_trans structure (1939 structures), 4.0% around the S0_cis structure (1776 geometries), 0.2% around the MECI_opt geometry (112 geometries) and 0.4% around the MECI_db structure (188 geometries), only visited slightly more frequently than the *ab initio* optimised structure.

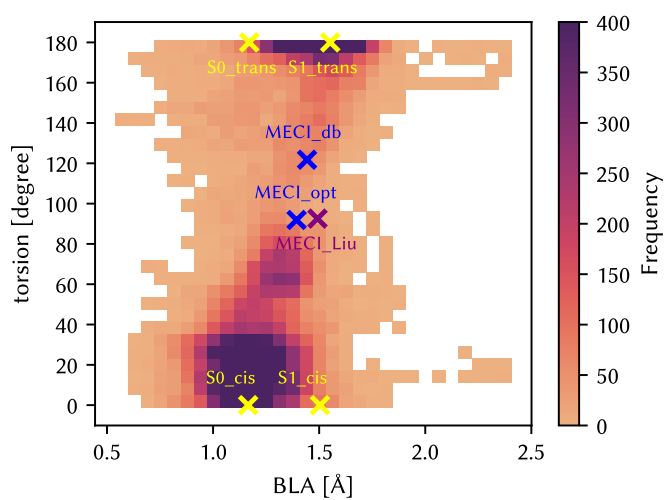

**Fig. 4** Histogram of the torsional angles (C2-C3-C4-C5) and bond alternation angles (BLA) calculated as BLA = $r_{C1-C2}$ - $r_{C2-C3}$ + $r_{C3-C4}$ $r_{C4-C5}$ +$r_{C5-N}$ for the 48,857 database structures. S0_trans and S0_cis represent the initial and final trans and cis- PSB3 structures on the ground electronic state found in the DD-vMCG/20GBFs dynamical simulation. MECI_db is the minimum energy conical intersections also found in the dynamics. S1_trans, S1_cis are the *ab initio* optimised trans and cis-PSB3 on the excited state, respectively. MECI_opt is the *ab initio* optimised minimum energy conical intersection. MECI_Liu corresponds to the conical intersection obtained by Liu *et al.*[25]. *Ab initio* optimisation calculations were done using a SA(2)-CAS(6,6)SCF/6-31G with Molpro program[35].

The excitation energies of the cis, trans isomers and MECI structures that we report in this work are in good agreement to those reported by other authors, with MECI energies laying at 2.87,[25] 2.89,[27] and 2.87 eV[26]. However, it is clear to us that the deactivation pathway in our case is involving an energy barrier. From Figure 4, the system is initially in S0_trans and gets trapped in the S1_trans minimum (almost 2,000 structures of the database seem to be at the torsional and BLA values of the S1_trans minimum geometry found *ab initio*). After that, the system slowly manages to leave that local minimum and continues evolving towards torsional angles of 90-120 degree and BLA values of 1.4, starting to transfer population to the groundstate at 170 fs. At 300 fs, the electronic population is still 60:40 (S1:S0), but the system seems to have relaxed to the S0_cis structure (1,800 structures) without passing by the S1_cis minimum geometry. This suggests a similar deactivation mechanism that other authors report through a similar conical intersection, but the system is trapped for a while at the more stable S1_trans minimum, and from there is able to relax through both the MECI_db minimum and the MECI_opt minimum.

The PES showing this energy barrier can be visualised in Figure 5. This energy barrier explains the delayed isomerisation observed in our simulations, as other studies report isomerisation timescales as short as 60 fs.

Additionally, in Figures S13 and S14 we show the distribution of energies of the database geometries along the torsion and the BLA dimensions. The MECI_db is clearly visible at 120 degree of torsion and a BLA of 1.4, whereas another region of almost degeneracy between the ground and the excited state can be observed at torsions of around 80 degrees, which could correspond to the MECI_opt, indicating two possible deactivation routes.

### 3.3 Dynamics on reduced dimensionality

To investigate the impact of including or neglecting degrees of freedom on the system's dynamics, we selectively froze several normal modes and re-ran the simulations. Initially, we performed a calculation across the full conformational space, encompassing 36 normal modes, using a nuclear basis of 16 frozen Gaussians. The results, depicted as the dark purple "36mode" solid line in Figure 6, are based on a database containing 45,000 records. This simulation demonstrated that the system relaxed to the ground state on a similar timescale (approximately 250 fs) as the previously reported calculations with 20 Gaussians. Subse-quently, reduced-dimensionality calculations were carried out, focusing on 13 (q4,q9,q15,q16,q19,q20,q22,q23,q27,q29,q30,q32,q36), 19 (q1-q4,q6-q9,q15,q16,q19-q27), and 27 modes (q1-q27). The calculations were run independently -not based on the database of the simulations in full dimensionality- and resulted in databases of 42,701 (13mode) and 44,929 (19 and 24mode), the latter being shared by both reduced-dimensionality calculations. None of these simulations successfully relaxed to locate the cis-isomer. Notably, the 27-mode calculation excluded only the CH stretch modes, highlighting the crucial role of these modes in the isomerisation mechanism.

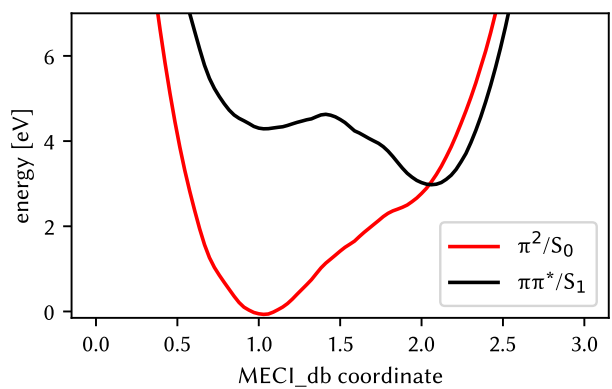

**Fig. 5** Energy profile for the two electronic states extracted from the database with 48,857 geometries. The x axis is the vector that connects the S0_trans (at 1.0 arbitrary units) and the MECI_db (at 2.0 arbitrary units) geometries.

## Conclusions

This study investigated the photochemical trans-cis isomerisation of the PSB3 retinal model using nonadiabatic quantum dynamics simulations based on the DD-vMCG method. The main findings can be summarized as follows: upon vertical excitation from the Franck-Condon region to the first electronic state ($\pi\pi^*/S_1$), the wavepacket initiates its dynamics, first being trapped in the minimum of the excited state with larger BLA values and then evolving toward the conical intersections (shorter BLA and 120/80 degree of torsion respectively) which facilitates population transfer back to the ground state ($\pi^2/S_0$). The isomerisation is driven by coupled vibrational modes, notably torsional and CC stretching modes. These modes guide the wavepacket through key regions of the PES associated with the isomerisation pathway. Based on reduced dimensionality results, we observed that CH stretches are crucial for a succesful isomerisation.

While our simulations reveal significant $S_1$-to-$S_0$ transfer only after 200 fs, mixed quantum-classical approaches such as Ab Initio Multiple Spawning (AIMS)[25] and Tully Surface Hopping[27] show nearly complete transfer within this timescale. The final cis-PSB3 geometry is achieved by 250–300 fs, with key signatures such as a torsional angle near 0 degree and bond-length alternation (BLA) values stabilizing at ≈1.4 Å, confirming successful isomerisation, as depicted in Figure 7.

In conclusion, in this work we tested the DD-vMCG method on the trans-PSB3 system, which has been widely studied using mixed quantum–classical trajectory methods. The tested method is able to predict the photoisomerisation of the retinal toy model, although discrepancies in timescales and pathway exploration between quantum and mixed quantum–classical methods are observed. These results are not entirely unexpected, since the DD-vMCG method -although still in its early stages of development- is a fully variational wavepacket approach which may yield different results than independent trajectory methods. Similar discrepancies have also been reported in previous quantum dynamics studies using grid-based MCTDH methods for molecules like ethylene[48,49] and 1,1-difluoroethylene.[50,51]. These findings reinforce the importance of continued investigation into the origins of such differences and how they relate to the regions of configurational space sampled by each method. Overall, we find these results encouraging and supportive of further development and application of *on-the-fly* quantum dynamical approaches to complex nonadiabatic processes.

## Author contributions

María Mallo: Formal analysis (equal); Methodology (equal); Writing – original draft (lead). Susana Gómez-Carrasco: Formal analysis (equal); Writing – review & editing (equal). Sandra Gómez: Formal analysis (equal); Methodology (equal); Funding

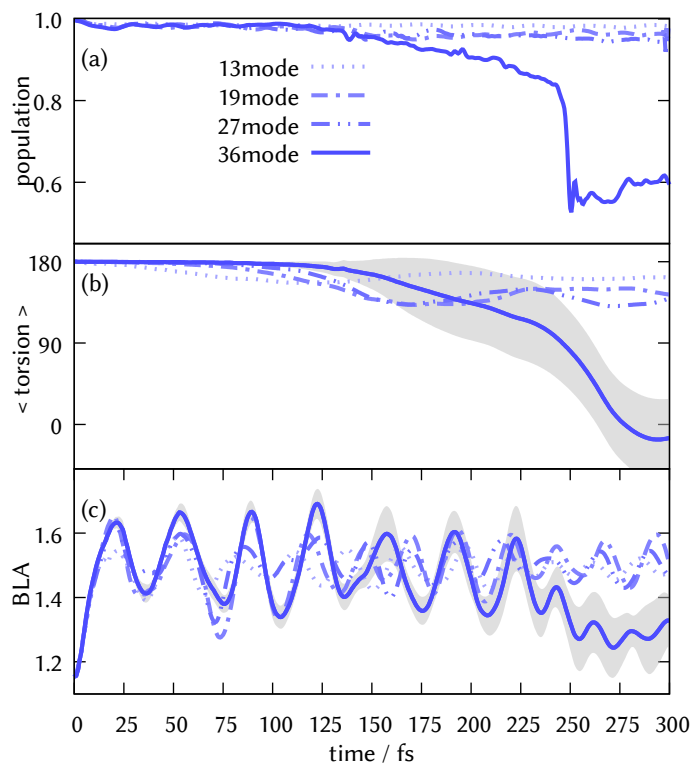

**Fig. 6** (a) Diabatic electronic population decay for the initially excited state ($\pi\pi^*/S_1$), (b) average torsional angle (C1-C2-C3-C4) in degrees and standard deviation of the gaussian centers in cartesian coordinates, (c) Evolution of bond length alternation (BLA) parameter in Å defined as $r_{C1-C2} - r_{C2-C3} + r_{C3-C4} - r_{C5-N}$. The dashed lines correspond to DD-vMCG simulations with 16GWPs with a reduced subset of normal modes, freezing some of the degrees of freedom. The solid line corresponds to the DD-vMCG/16GWPs simulation in full dimensionality.

acquisition (lead); Methodology (equal); Project administration (lead); Supervision (lead); Writing – review & editing (equal).

## Conflicts of interest

There are no conflicts of interest to declare.

## Data availability

All data necessary to reproduce the results presented in this article are available at Zenodo at https://doi.org/10.5281/zenodo.15582165. The dataset includes the database created during the dynamics in .sql format for the 20GWP and 16GWP cases and the Supporting Information as a PDF file, which includes optimised structures for the $S_0$, $S_1$ states and the conical intersection (Section S1), vibrational frequencies (Section S2), convergence of diabatic electronic populations (Section S3), tiem evolution of each normal mode (Section S4), the evolution of dihedral angles for each frozen Gaussian with its weight in the expansion (Section S5), and the distribution of 48,857 database geometries along the torsional angle and the BLA coordinate and their energies (Section S6).

## Acknowledgements

SG acknowledges NextGenerationEU funds (María Zambrano Grant for the attraction of international talent) and the grant "Programa Propio C1" from Universidad de Salamanca. The authors thank the funding by Spanish Ministry of Science and Innovation (MCIN/AEI/10.13039/501100011033) grant No. PID2020-113147GA-I00, the EPSRC under the COSMOS programme grant (EP/X026973/1) and the COST action CA21101" Confined molecular systems: from a new generation of materials to the stars (COSY)" supported by COST (European Cooperation in Science and Technology). The authors thank Graham A. Worth and Olivia Bennett for their valuable input and feedback.

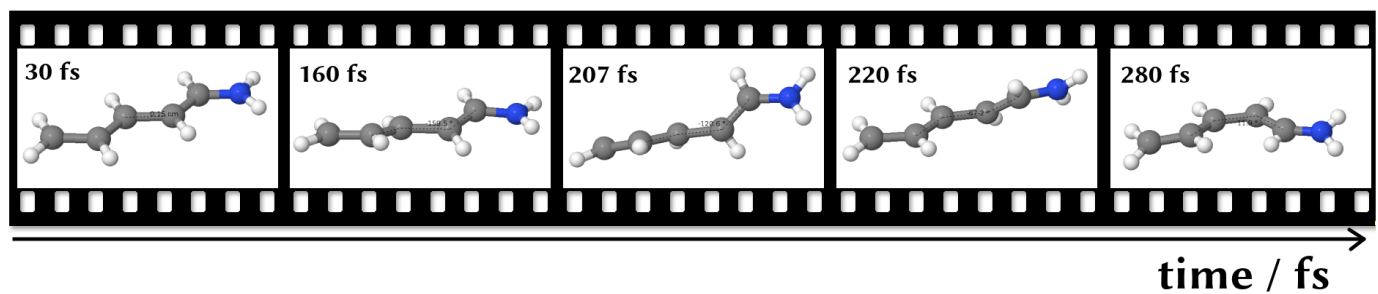

time / fs

**Fig. 7** Average structure of PSB3 (averaged over the 20 gaussian centers with a weight proportional to the amplitude of each gaussian in the linear combination -$A_i$ in equation 1-) for the DD-vMCG/20GWP propagation, showing how trans-PSB3 isomerises to cis-PSB3 in 300 fs.

## Notes and references